# Realization of a Wigner-Mott insulating state in 6*R*-$TaS_2$ superconductor

Hongqin Xiao[1,2#], Geng Li[1,2,3#,*], Yuxuan He[1,2#], Ke Zhu[1,2#], Yuhan Ye[1,2,3], Yumeng Li[1,2], Lijing Huang[1,2], Pucen Xiong[1,2], Haitao Yang[1,2,3], Ziqiang Wang[4] & Hong-Jun Gao[1,2,3*]

[1] *Beijing National Center for Condensed Matter Physics and Institute of Physics, Chinese Academy of Sciences, Beijing 100190, China*

[2] *School of Physical Sciences, University of Chinese Academy of Sciences, Beijing 100190, PR China*

[3] *Hefei National Laboratory, Hefei 230088, China*

[4] *Department of Physics, Boston College, Chestnut Hill, MA 02467, USA*

**Wigner-Mott insulating states represent a paradigmatic manifestation of strong electronic correlations, in which long-range Coulomb interactions drive spontaneous charge ordering and enable Mott localization at fractional electronic fillings[1-5]. Such states have been theoretically proposed to arise from the cooperative interplay between onsite and inter-site Coulomb interactions[1,2]. However, experimental realizations of the simultaneous microscopic observation of interaction-driven charge order and genuine Mott localization, which are the defining hallmarks of a Wigner-Mott insulator, have remained elusive. Here we report the observation of a Wigner-Mott insulating state in 6*R*-$TaS_2$ using scanning tunneling microscopy. By locally injecting electrons into the depleted 1*T* layer, we induce distinct Star-of-David charge-ordered superstructures and realize a cascade of insulating phases. In particular, a $\sqrt{3}\times\sqrt{3}$ charge-ordered superstructure at one-third filling hosts a robust Mott gap despite fractional filling. The spontaneous relaxation from excited states back to the ground state demonstrates that this Wigner-Mott phase is stabilized by the cooperative effects of onsite and inter-site Coulomb interactions. Our results provide direct microscopic evidence for a Wigner-Mott mechanism and establish 6*R*-$TaS_2$ as a platform for the controlled realization and investigation of Wigner-Mott insulating states.**

[#]These authors contributed equally to this work.

[*]Correspondence to: hjgao@iphy.ac.cn; gengli.iop@iphy.ac.cn.

The Mott insulating state, in which strong onsite Coulomb repulsion $U$ prohibits double electron occupancy and forces electrons to localize on individual lattice sites, forms the foundation for a wide range of correlated quantum phases. Doping a Mott insulator away from integer-filling leads to exotic quantum states, such as unconventional superconductivity[6-8], spin density wave[9,10], and other exotic collective states[3,11-13]. Beyond this conventional paradigm, it has been theoretically proposed that Mott localization may also emerge at fractional fillings when inter-site Coulomb interactions drive spontaneous charge ordering, effectively generating an emergent superlattice with an odd number of electrons per unit cell[1,2]. In this scenario, commonly referred to as a Wigner-Mott insulator, charge ordering driven by nonlocal interactions effectively restores integer filling on the emergent supercell, enabling Mott localization far away from integer filling[1,2]. While fractional-filling insulating states have been extensively explored in moiré superlattices[3-5,14], experimental signatures of Wigner-Mott physics have so far remained indirect. In particular, the simultaneous microscopic observation of interaction-driven charge order and genuine Mott localization characterized by well-defined Hubbard bands and a hard Mott gap within the same phase has yet to be achieved.

6*R*-$TaS_2$ provides a unique layered platform for investigating strong electronic correlations in reduced dimensions[15-19]. In this van der Waals crystal, 1*T* and 1*H* layers of $TaS_2$ are alternately stacked along the *c*-axis, forming electronically distinct layers that are weakly coupled through interlayer charge transfer[20]. This natural heterostructure enables the coexistence and interplay of localized and itinerant electronic states, creating a favorable environment for correlation-driven phenomena. Indeed, a variety of exotic quantum states have been reported in related $TaS_2$-based layered systems, including artificial heavy-fermion behavior arising from Kondo coupling in monolayer 1*T*/1*H* heterostructures[21], as well as topological superconductivity in stacked 1*T*/1*H* compounds[22,23]. More recently, transport measurements[16] have suggested the presence of a hidden ordered state that breaks time-reversal symmetry in 6*R*-$TaS_2$. The coexistence and competition among these correlated electronic phases highlight 6*R*-$TaS_2$ as a promising platform for exploring interaction-driven insulating states, including Mott and Wigner-Mott physics.

In this Letter, we report the real-space visualization of Wigner-Mott insulating states in the layered superconductor 6*R*-$TaS_2$. Using scanning tunneling microscopy and spectroscopy (STM/STS), we locally inject electrons into the 1*T* layer and controllably tune the electron filling, thereby generating a cascade of emergent insulating phases. At the microscopic level, we resolve distinct regimes of conventional Mott localization and Wigner-Mott localization, distinguished by the absence or presence of spontaneous charge ordering.

6*R*-$\mathrm{TaS_2}$ crystallizes in a space group of R3m with a stacking sequence of alternating 1*T* and 1*H* layers (Fig. 1a), and lattice parameters $a_0 = b_0 = 3.3$ Å and $c_0 = 35.8$ Å (Extended Data Fig. 1). These layers are coupled through van der Waals forces, with each 1*T* (1*H*) layer sliding one-third of the lattice constant with respect to the adjacent layer of the same type. Upon cleavage, the surface will reveal either the 1*T* or 1*H* configuration (Fig. 1b,c). The crystal undergoes a commensurate CDW transition at ~303 K, and exhibits superconductivity at $T_c$~3.3 K (Extended Data Fig. 1c,d). Below the CDW transition temperature, the 1*T* layer undergoes a periodic lattice distortion (Fig. 1b), with 13 Ta atoms forming a characteristic Star-of-David (SoD) structure (Fig. 1b). Consequently, a $\sqrt{13}\times\sqrt{13}$R13.9° CDW pattern emerges (Fig. 1d-f).

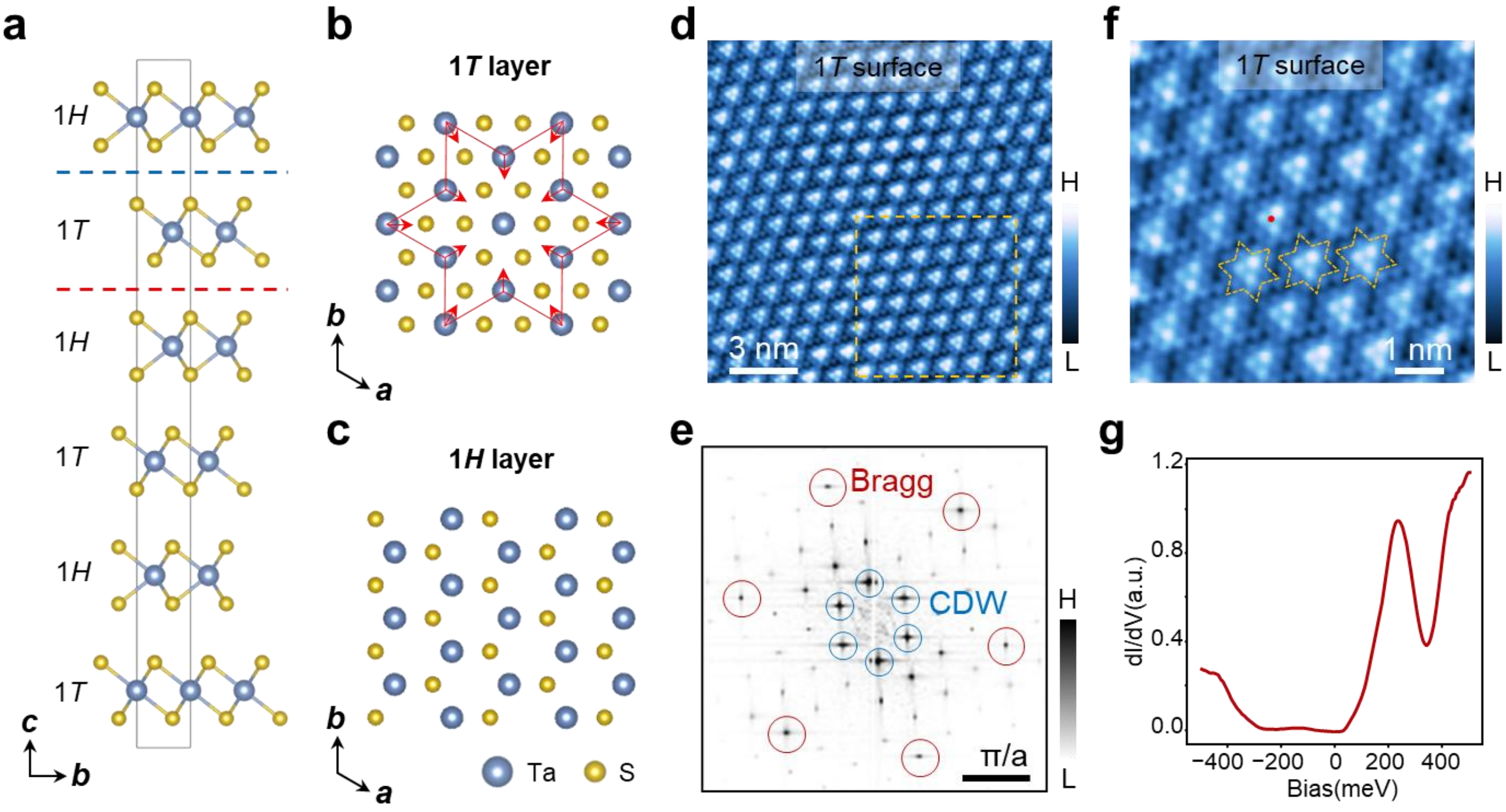

**Fig. 1 | Crystal structure and STM/S of the cleaved surfaces of 6*R*-$\mathrm{TaS_2}$. a,** Side view of the crystal structure of 6*R*-$\mathrm{TaS_2}$, showing stacking of 1*T* and 1*H* layers along the *c* axis. **b,c,** Top view of the 1*T* and 1*H* layers, respectively. **d,** STM image of the 1*T* surface, showing the $\sqrt{13}\times\sqrt{13}$ SoD pattern (scanning settings: $V_s$=500 mV, $I_t$=1 nA). **e,** FT image of **d**, showing clearly the $\sqrt{13}\times\sqrt{13}$ CDW spots. **f,** Atomically-resolved STM image of the yellow square in **d** (scanning settings: $V_s$=500 mV, $I_t$=1 nA). **g,** d*I*/d*V* spectrum taken at the center of a SoD (red dot in **f**).

In a pristine 1*T* layer, each SoD hosts one localized unpaired 5*d* electron from the center Ta atom, leading to a Mott insulating state. However, in the stacked 1*T*/1*H* heterostructure, substantial interlayer charge transfer depletes the $1T$ layer of electrons, driving it far below integer filling of one-electron per SoD, while preserving the underlying SoD unit cell, as have been observed in $4Hb$-$TaS_2$ (Ref. [23,24]) and the $1T/1H$ monolayer heterostructure[20,21,25-27]. In $6R$-$TaS_2$, similar charge transfer from the $1T$ to the $1H$ layers depletes electrons from the SoDs, resulting in a heavily hole-doped Mott insulating phase with dilute electron densities, as can be seen from the location of the Fermi level near the bottom of the lower Hubbard band (LHB) (Fig. 1g). As a result, the $1T$ layer in $6R$-$TaS_2$ realizes a highly correlated, low-density metallic system that is qualitatively distinct from both the stoichiometric Mott state of bulk $1T$-$TaS_2$ and a conventionally doped Mott insulator. In this regime, long-range Coulomb interactions are expected to play a dominant role, opening the possibility for electrons to self-organize into emergent charge-ordered superlattices and thereby re-establish Mott localization at fractional fillings described by the Wigner-Mott scenario. In contrast, the $1H$ layer develops a $3\times3$ CDW phase (Extended Data Fig. 2a,b). At temperatures below $T_c$, superconductivity develops exclusively in the $1H$ layers, with a V-shaped nodal gap[16] of ~0.46 meV (Extended Data Fig. 2c). This gives $2\Delta/k_B T_c \sim 3.23$, revealing a weak coupling superconducting state[23,28].

We develop a technique to tune the electron filling in the SoDs of the surface $1T$ layer. The voltage pulses applied via an STM tip can induce a transient current flow, injecting electrons[29] into the $1T$ layer (Fig. 2a). These injected electrons become trapped in the SoDs, giving rise to a mosaic pattern that appears as scattered bright SoDs (Fig. 2b,c). Notably, this mosaic pattern emerges only when the applied voltage exceeds 1.2 V, indicating a threshold energy for the electron injection process. With the tip further approaching the substrate and performing slow, repeated scans in this region, the electron-injected-SoDs (EI-SoDs) are gradually depleted via tunneling events, ultimately leading to the recovery of the original $\sqrt{13}\times\sqrt{13}$ CDW (Fig. 2d). As a result, the creation and removal of the EI-SoDs is reversible.

Under high voltage pulses, a large-scale mosaic pattern is observed (Fig. 2e). The EI-SoDs could form short-range ordered superstructures[30] due to inter-site Coulomb interactions. Depending on the filling level of electrons injected by the STM tip, we observe two distinct ordered superstructures of EI-SoD clusters: a $1\times1$ lattice (Fig. 2f) and a $\sqrt{3}\times\sqrt{3}$ lattice (Fig. 2g). In the $1\times1$ phase, each EI-SoD hosts a single injected electron, corresponding to an integer filling of one electron per SoD. By contrast, the $\sqrt{3}\times\sqrt{3}$ superstructure corresponds to a fractional filling of one electron per three SoDs (Extended Data Fig. 3). These ordered domains typically extend over lateral length scales of approximately 20-30 nm. Corresponding FT image shows clearly the coexistence of the $\sqrt{3}\times\sqrt{3}$ superstructure and the $\sqrt{13}\times\sqrt{13}$ CDW (Fig. 2g,h).

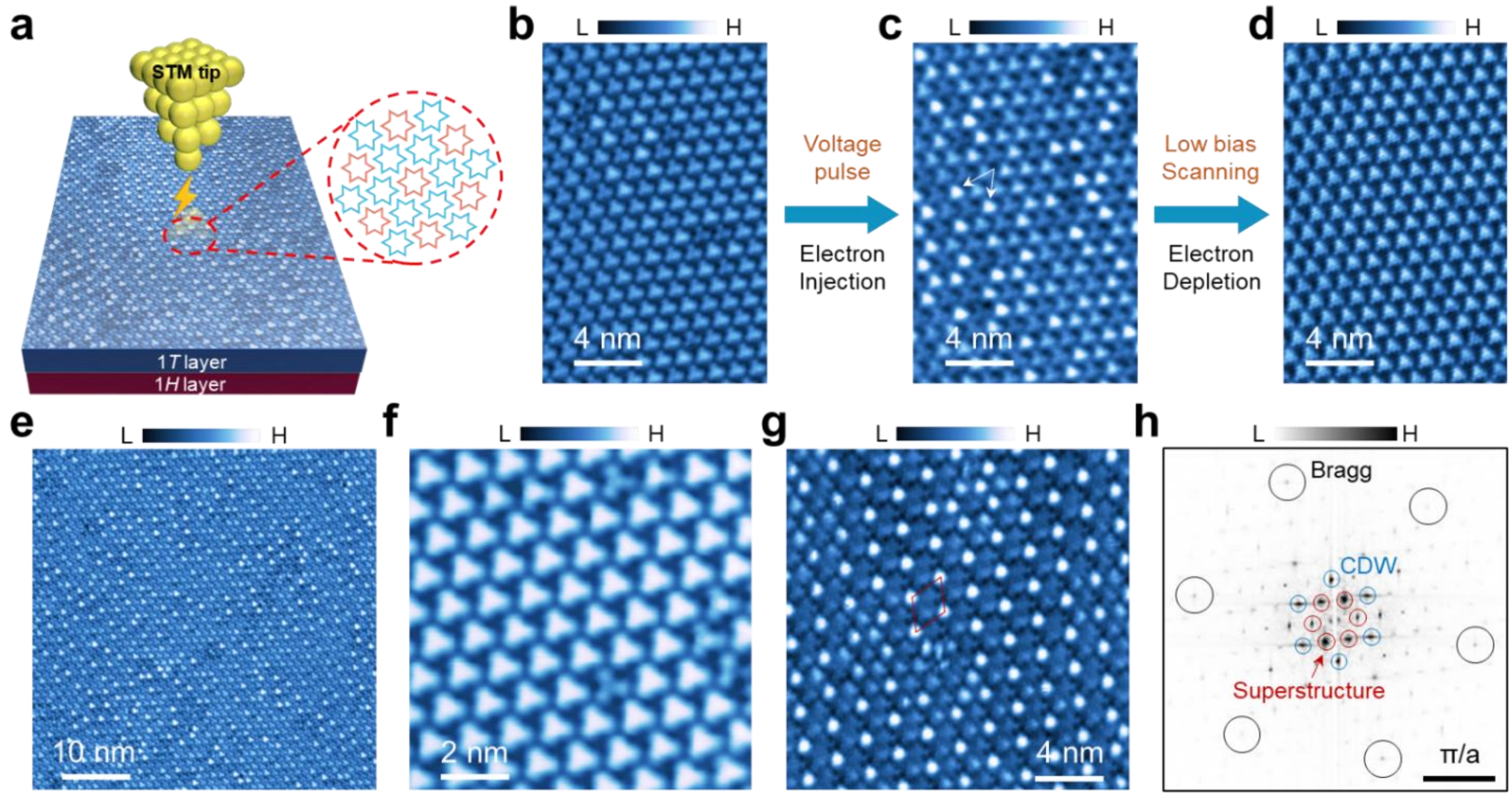

**Fig. 2 | Creation and depletion of electron-injected SoDs on the surface of 6*R*-$TaS_2$ 1*T* layer. a,** Schematic of the electron injection on the surface with STM voltage pulses. **b,** STM image of the 1*T* surface showing the periodic SoD lattice ($V_s$=-1000 mV, $I_t$=1 nA). **c,** STM image of **b** after applying STM voltage pulses, showing creation of EI-SoDs ($V_s$=-1000 mV, $I_t$=1 nA). **d,** STM image of **c** after continuous low-bias scanning, showing the depletion of the EI-SoDs ($V_s$=-1000 mV, $I_t$=1 nA). **e,** Large-scale STM image of the mosaic superstructure ($V_s$=-500 mV, $I_t$=1 nA), showing the coexistence of normal SoDs and EI-SoDs. **f,g,** Typical STM images in different regions of the voltage pulsed surface, showing the formation of ordered $1\times1$ (**f**) and $\sqrt{3}\times\sqrt{3}$ (**g**) superstructures, respectively ($V_s$=-500 mV, $I_t$=1 nA). **h,** FT image of **g**, showing both the CDW and $\sqrt{3}\times\sqrt{3}$ superstructure on the EI-SoD surface.

To elucidate the nature of electron correlations in these phases, we performed spatially resolved scanning tunnelling spectroscopy (STS). In the $1\times1$ phase, a robust hard gap of ~330 meV is consistently observed (Fig. 3a-c), accompanied by well-defined lower and upper Hubbard bands (LHB and UHB) that closely resemble those of bulk 1*T*-$TaS_2$ crystal (Extended Data Fig. 4)[31,32]. These spectroscopic signatures indicate the emergence of a Mott insulating state, consistent with the localization of one electron on each depleted SoD cluster. Effectively, we utilized nanoscale manipulation to recover the pristine bulk 1*T*-$TaS_2$ Mott insulating state within the electron-depleted 1*T* layer in 6*R*-$TaS_2$. In the $\sqrt{3}\times\sqrt{3}$ superstructure, we observe qualitatively similar Hubbard-band features, although both the LHB and UHB are shifted slightly towards higher energies (Fig. 3d-f,

Extended Data Fig. 5). Notably, this insulating state develops at one-third filling relative to the underlying SoD lattice, implying that Mott localization is stabilized not by half-filling of the original SoD lattice but by interaction-driven charge ordering. This behavior is consistent with a Wigner-Mott scenario, in which inter-site Coulomb interactions induce an emergent superlattice that restores an odd number of electrons per $\sqrt{3}\times\sqrt{3}$ supercell.

To verify the presence of charge ordering in the $\sqrt{3}\times\sqrt{3}$ superstructure, which is an essential requirement of the Wigner-Mott insulating state, we perform detailed STM measurements. Similar superstructure is observed in transition-metal dichalcogenides via physical adsorption of alkali atoms[33,34], chemical substitution[35] and chemical intercalation[36]. Importantly, this $\sqrt{3}\times\sqrt{3}$ charge-ordered superstructure exhibits a pronounced contrast reversal under opposite bias polarities. The EI-SoD clusters appear as bright protrusions at a sample bias of -400 mV (Fig. 3g), whereas they become dim sites at 200 mV (Fig. 3h). This bias-dependent contrast reversal reflects a modulation of the local electronic density of states, confirming the electronic origin of the superstructure. Together with the observation of a robust Mott gap at one-third filling, these results provide compelling microscopic evidence that the $\sqrt{3}\times\sqrt{3}$ phase realizes a Wigner-Mott insulating state.

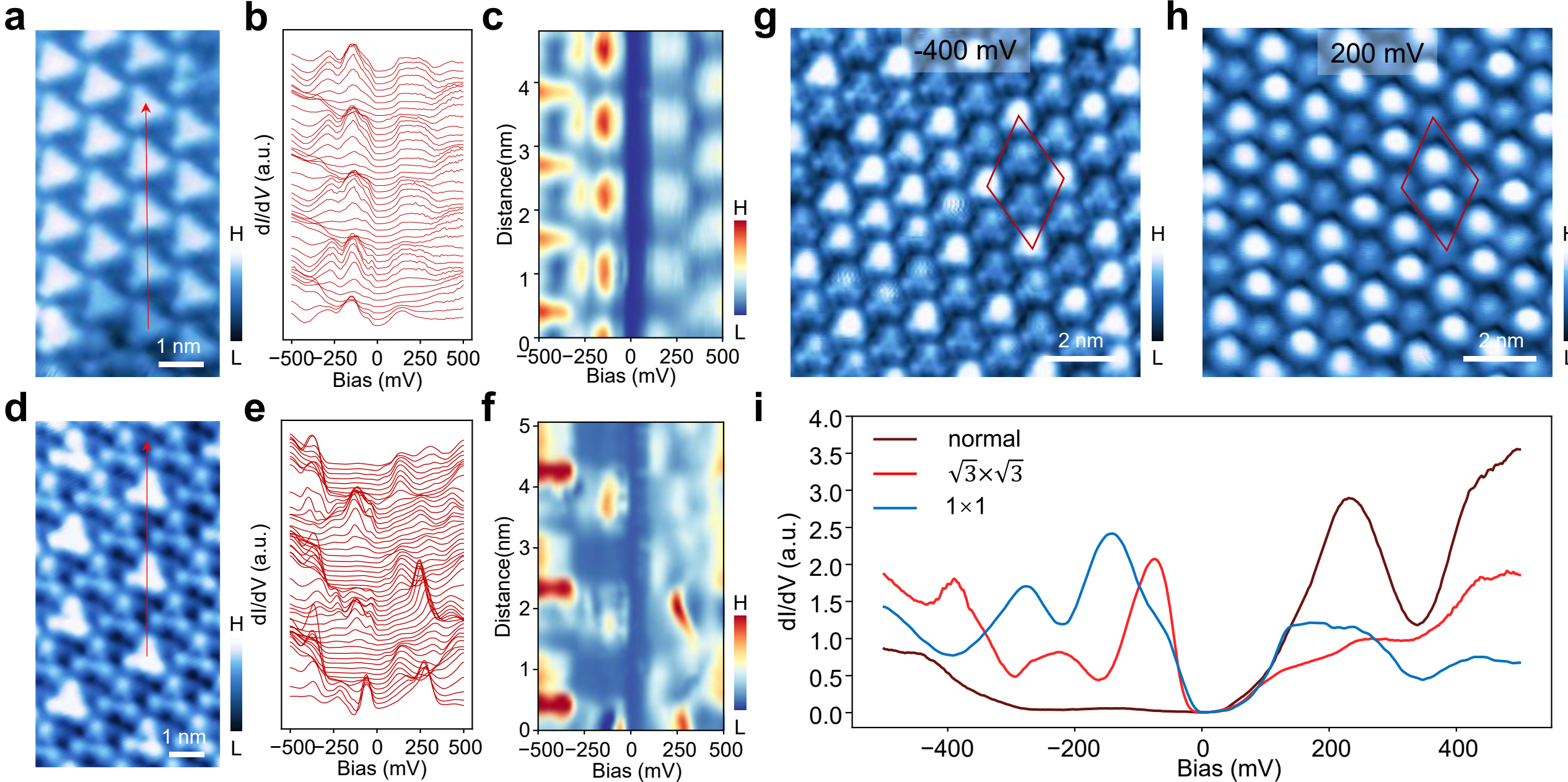

**Fig. 3 | STM/S of the Mott and Wigner-Mott insulating states in the ordered superstructures. a,** STM topography of a representative 1×1 superstructure ($V_s$ =-500 mV, $I_t$=1 nA). **b,c,** Waterfall plot and intensity map of the d$I$/d$V$ spectra along the red arrow in **a**. **d,** STM topography of a representative $\sqrt{3}\times\sqrt{3}$ superstructure ($V_s$ =-500 mV, $I_t$=1 nA). **e,f,** Waterfall plot and intensity map of the d$I$/d$V$ spectra along the red

arrow in **d**. **g,h,** STM images of the same area at -400 mV (**g**) and 200 mV (**h**), showing contrast reversal of the $\sqrt{3}\times\sqrt{3}$ charge order ($I_t$=1 nA). The red rhombuses mark the same unit cell of the superstructure at different bias voltages. **i,** Averaged d$I$/d$V$ spectra taken at the centers of SoDs in the normal region (brown), and in the electron-injected SoDs of the $\sqrt{3}\times\sqrt{3}$ (red) and $1\times1$ (blue) superstructures.

We next compare the d$I$/d$V$ spectra across the distinct superstructures realized at different electron filling levels . In the original SoD lattice, the spectrum displays a pronounced peak at ~230 mV, which is attributed to the LHB[32,37,38] of a hole-doped 2D Mott insulator. In this regime, the Fermi level lies at the bottom edge of the LHB, placing the system in a dilute metallic state with a low electron density per SoD. Upon progressive electron injection from the tip, the electronic filling is continuously increased, which is directly reflected in the systematic evolution of the LHB (Fig. 3i). As the filling increases, the Fermi level moves upward within the LHB, providing a transparent spectroscopic measure of the changing carrier density.

At one-third filling, a $\sqrt{3}\times\sqrt{3}$ superstructure of EI-SoDs emerges. In this phase, the Fermi level moves within the band renormalized by the charge order, opening a hard Mott gap of approximately 330 meV (Fig. 3i). Remarkably, despite the fractional filling, the system exhibits a fully developed Mott gap accompanied by a spontaneously formed charge-ordered superlattice, signaling the emergence of a Wigner-Mott insulating ground state. With further electron injection, the system reaches the $1\times1$ superstructure, corresponding to one electron per SoD. In this integer-filled phase, the LHB shifts to lower energy (−140 meV), and a robust Mott gap persists between the LHB and UHB, characteristic of a conventional Mott insulator (Fig. 3i and Extended Data Fig. 6). These spectroscopic results establish a filling-controlled cascade of correlated electronic states, evolving continuously from a dilute metal to a fractional-filling Wigner-Mott insulator and finally to an integer-filling Mott insulator.

Having established the distinct correlated ground states across different filling regimes, we next investigate the excited states of the Wigner-Mott insulating phase. Remarkably, in the $\sqrt{3}\times\sqrt{3}$ phase, we observe pronounced ring-like features in spatially resolved d$I$/d$V$ maps, which provide direct access to the elementary excitations of the Wigner-Mott insulating state (Extended Data Fig. 7). These ring states emerge periodically across the superstructure and display a striking particle-hole asymmetry: under positive bias voltages, the rings form an ordered triangular lattice centered on the EI-SoD sites (Fig. 4a,b), whereas under negative bias voltages they reorganize into a honeycomb lattice centered on the remaining, nominally unoccupied SoDs (Fig. 4c). The ring diameters vary spatially and increase linearly with the applied bias magnitude for both polarities (Fig. 4d-i and Extended Data Fig. 8), indicating a well-defined excitation energy scale. We note that the ring states have been theoretically predicted in topological insulators and Mott insulators[39,40], where local potentials induced by defects create bound states through doublon-holon coupling.

Importantly, these ring states represent localized, intra-supercell excitations of the $\sqrt{3}\times\sqrt{3}$ Wigner-Mott superlattice, with energies above the Mott-Hubbard gap. Microscopically, they correspond to excited configurations in which an electron is transiently displaced within a single $\sqrt{3}\times\sqrt{3}$ supercell, while the overall charge order and global filling of the Wigner-Mott state remain conserved. Such intra-unit-cell excitations are a defining feature of Wigner-Mott insulators. While the insulating behavior in the $1\times1$ phase is governed predominantly by the onsite Coulomb interaction $U$, the $\sqrt{3}\times\sqrt{3}$ phase at one-third filling requires the cooperative interplay of $U$ and inter-site Coulomb interaction $V$, which stabilizes both the charge-ordered superlattice and the Mott gap[4,5,41].

In STM measurements, the tip acts as a highly localized electrostatic perturbation, effectively serving as a local top gate that promotes these intra-supercell excitations. In the dilute limit, this process bears a close resemblance to the well-known tip-induced charging and discharging phenomena commonly observed in STM studies[42-45]. Within this classical picture, when the tip is positioned above an EI-SoD, the strong electric field can transiently remove an electron under positive sample bias, producing ring-like features in d$I$/d$V$ maps accompanied by sharp spectral resonances. Conversely, under negative bias, electrons can be transferred from EI-SoDs to initially depleted SoDs, giving rise to complementary ring patterns. This analogy provides an intuitive interpretation of the bias polarity dependence and the characteristic ring geometry observed in our measurements. However, unlike a classical Wigner crystal, the $\sqrt{3}\times\sqrt{3}$ phase is a Wigner-Mott insulator, and the classical charging-discharging picture alone is insufficient to capture the full nature of the observed excitations. Crucially, the Wigner-Mott insulating state is realized only at the specific fractional filling of one electron per $\sqrt{3}\times\sqrt{3}$ supercell, where strong correlations open a robust Mott gap while long-range Coulomb interactions stabilize the charge-ordered superlattice. The ring states observed here therefore reflect many-body excitations of a correlated ground state, rather than simple single-electron charging events.

Consistent with this interpretation, the ring-like features emerge exclusively within the $\sqrt{3}\times\sqrt{3}$ superstructure. By deliberately identifying a boundary between the $\sqrt{3}\times\sqrt{3}$ phase and the surrounding depleted SoD region, we find that the ring states are strictly confined to the charge-ordered superstructure and are entirely absent in the normal region (Extended Data Fig. 9). Moreover, no such ring features are detected in either the integer-filled $1\times1$ Mott phase or the bulk 1$T$-$TaS_2$ crystal, ruling out charge exchange with the underlying 1$H$ layer within the experimental energy window. These observations demonstrate that the ring states are intrinsic excitations of the $\sqrt{3}\times\sqrt{3}$ Wigner-Mott phase.

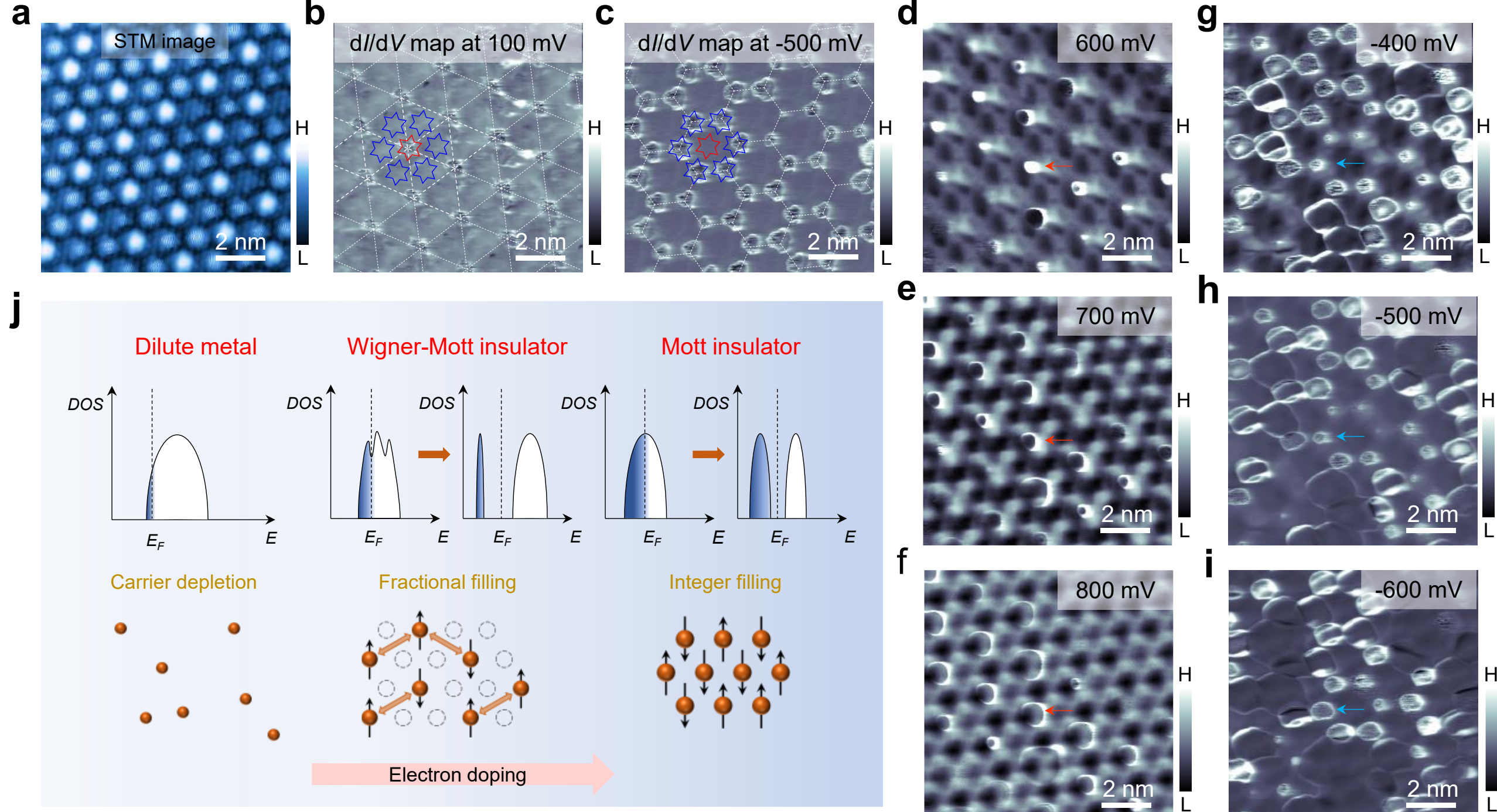

**Fig. 4 | STM/S of the $\sqrt{3}\times\sqrt{3}$ Wigner-Mott insulating phase on the 1*T* surface. a,** STM topography of a representative $\sqrt{3}\times\sqrt{3}$ superstructure ($V_s$ =-500 mV, $I_t$=1 nA). **b,** d$I$/d$V$ map of the region in **a** at 100 mV, showing the formation of triangular lattice of rings centered at the electron-injected SoDs (red)**. c,** d$I$/d$V$ map of the region in **a** at -500 mV, showing formation of honeycomb lattice of rings centered at the remaining SoDs (blue)**. d-f,** d$I$/d$V$ maps of a $\sqrt{3}\times\sqrt{3}$ superstructure at 600, 700 and 800 mV, respectively, showing spatial dispersion of the discharging rings. **g-i,** d$I$/d$V$ maps of a $\sqrt{3}\times\sqrt{3}$ superstructure at -400, -500 and -600 mV, respectively, showing similar spatial dispersion of the charging rings. **j,** Schematic of the ground states of the dilute metal, Wigner-Mott insulator and Mott insulator at different doping level.

To further elucidate the role of inter-site interactions, we perform repeated STM scans at bias voltages exceeding the excitation threshold. Although the tip can transiently drive the system into excited configurations, the electrons consistently relax back to their original EI-SoD sites once the tip is removed, and the $\sqrt{3}\times\sqrt{3}$ charge-ordered pattern remains unchanged over repeated measurements (Fig. 4d-i and Extended Data Fig. 10). This reversible relaxation behavior indicates that the $\sqrt{3}\times\sqrt{3}$ Wigner-Mott state represents a global energy minimum of the system. The spontaneous recovery of the charge order against local perturbations provides direct evidence that inter-site Coulomb interactions $V$ actively stabilize the superlattice against the tip perturbation.

Our results establish a real-space realization of a Wigner-Mott insulating state, in which Mott localization and interaction-driven charge ordering coexist and mutually reinforce each other (Fig. 4j). In contrast to conventional Mott insulators governed predominantly by onsite $U$, and to Wigner crystals stabilized solely by long-range $V$, the $\sqrt{3}\times\sqrt{3}$ phase in 6$R$-$TaS_2$ exhibits a robust Hubbard gap that persists at fractional filling, and a spontaneous charge-ordered superlattice emerging from inter-site Coulomb interactions (Fig. 4j). The charge-ordered pattern is not only statically resolved but also dynamically stabilized, as evidenced by the reversible recovery of the superlattice following repeated STM/S measurements. This self-restoring behavior demonstrates that the $\sqrt{3}\times\sqrt{3}$ charge-ordered lattice represents a global energy minimum protected by inter-site Coulomb interactions and is robust against external perturbations. These results thus provide direct microscopic evidence for the cooperative roles of $U$ and $V$ in stabilizing a Wigner-Mott insulator.

In summary, we have realized and directly visualized a Wigner-Mott insulating state on the 1$T$-terminated surface of 6$R$-$TaS_2$. We achieve microscopic control of electron filling and uncover a $\sqrt{3}\times\sqrt{3}$ insulating phase characterized by a spontaneously formed charge-ordered superlattice and a robust Mott gap at fractional filling. The coexistence of real-space charge order and genuine Mott localization provides direct evidence for a Wigner-Mott insulating state, in which both onsite Coulomb repulsion $U$ and inter-site interaction $V$ play essential roles. Our results establish 6$R$-$TaS_2$ as a unique platform for the real-space investigation of Wigner-Mott physics and its elementary excitations.

# Methods

## Single-crystal growth

The 6*R*-$TaS_2$ single crystals were grown using chemical vapor transport techniques, with $NH_4Cl$ as the transport agent. A stoichiometric amount of high-purity tantalum powder (Ta, 99.99%) and sulfur lumps, along with 50 mg of ammonium chloride ($NH_4Cl$), were placed in a silica ampule and sealed under vacuum ($10^{-3}$ Torr). The ampule was then placed in a two-zone horizontal furnace, with the source end positioned in the hot zone and the growth end in the cold zone. The furnace was maintained at constant temperatures of 1050 °C and 950 °C for 150 hours. After the growth, the system was allowed to cool naturally to room temperature. High-quality 6*R*-$TaS_2$ single crystals were obtained from the cold end of the ampule.

## STM/S experiments

A 6*R*-$TaS_2$ crystal was mounted onto a STM sample holder in a glove box and then transferred to an ultra-high vacuum chamber. The crystal was cleaved *in-situ* at a temperature of 80 K. After cleavage, the sample was immediately transferred to the STM scanner. The STM/S measurements were conducted in an ultra-low temperature STM system. Tungsten tips were etched chemically and calibrated on Au(111) surfaces before use. The $dI/dV$ spectra and maps were obtained using a standard lock-in technique with a modulation voltage of 5 mV at 973.0 Hz. All the images, spectra, and $dI/dV$ maps were taken at 40 mK.

**Acknowledgements:** The work is supported by the National Key Research and Development Projects of China (2024YFA1207700, 2022YFA1204100), National Natural Science Foundation of China (62488201), the CAS Project for Young Scientists in Basic Research (YSBR-003), the Youth Innovation Promotion Association (2023005), and the Quantum Science and Technology-National Science and Technology Major Project (2021ZD0302700). Z.W. is supported by the US DOE, Basic Energy Sciences Grant No. DE-FG02-99ER45747.

**Author Contributions:** H.-J.G. supervised the project. H.-J.G. and G.L. designed the experiments. K.Z. and H.Y. prepared samples. H.X., G.L., Y.H. and Y.Y. performed STM experiments with guidance of H.-J.G. and Z.W. provided theoretical consideration. G.L., H.X., Z.W. and H.-J.G. analysed the data and wrote the manuscript. All of the authors participated in analysing experimental data, plotting figures, and writing the manuscript.

**Data availability:** Data measured or analysed during this study are available from the corresponding author on reasonable request. Source data are provided with this paper.

**Competing Interests:** The authors declare that they have no competing interests.